\documentclass[reqno]{amsart}
\usepackage{graphicx}  
\usepackage[margin=1in,tmargin=1.3in]{geometry}
\usepackage{url}        
\usepackage{amsmath}
\usepackage{amssymb}
\usepackage{algorithm}
\usepackage{subcaption}
\usepackage{algpseudocode}
\usepackage{natbib}
\usepackage[usenames, dvipsnames]{color}
\setcitestyle{round,aysep={},yysep={,}}

\begin{document}
\title{Buildup of Speaking Skills in an Online Learning Community: A Network-Analytic Exploration}

\author[]{Rasoul Shafipour, Raiyan Abdul Baten, Md Kamrul Hasan, Gourab Ghoshal, Gonzalo Mateos, Mohammed Ehsan Hoque}

\begin{abstract}

Studies in learning communities have consistently found evidence that peer-interactions contribute to students' performance outcomes. A particularly important competence in the modern context is the ability to communicate ideas effectively. One metric of this is speaking, which is an important skill in professional and casual settings. In this study, we explore peer-interaction effects in online networks on speaking skill development. In particular, we present an evidence for gradual buildup of skills in a small-group setting that has not been reported in the literature. Evaluating the development of such skills requires studying objective evidence, for which purpose, we introduce a novel dataset of six online communities consisting of $158$ participants focusing on improving their speaking skills. They video-record speeches for $5$ prompts in $10$ days and exchange comments and performance-ratings with their peers. We ask (i) whether the participants' ratings are affected by their interaction patterns with peers, and (ii) whether there is any gradual buildup of speaking skills in the communities towards homogeneity. To analyze the data, we employ tools from the emerging field of Graph Signal Processing (GSP). GSP enjoys a distinction from Social Network Analysis in that the latter is concerned primarily with the connection structures of graphs, while the former studies signals on top of graphs. We study the performance ratings of the participants as \textit{graph signals} atop underlying interaction topologies. Total variation analysis of the graph signals show that the participants' rating differences decrease with time (slope$=-0.04$, $p<0.01$), while average ratings increase (slope$=0.07$, $p<0.05$)---thereby gradually building up the ratings towards community-wide homogeneity. We provide evidence for peer-influence through a prediction formulation. Our consensus-based prediction model outperforms baseline network-agnostic regression models by about $23\%$ in predicting performance ratings. This, in turn, shows that participants' ratings are affected by their peers' ratings and the associated interaction patterns, corroborating previous findings. Then, we formulate a consensus-based diffusion model that captures these observations of peer-influence from our analyses. We anticipate that this study will open up future avenues for a broader exploration of peer-influenced skill development mechanisms, and potentially help design innovative interventions in small-groups to maximize peer-effects.

\end{abstract}

\flushbottom
\maketitle
\newcommand{\GG}{\fbox{\textsl{Gourab Ghoshal}}}
\thispagestyle{empty}

\section{Introduction}

The idea that learners actively construct knowledge as they build mental frameworks to make sense of their environments has been widely studied under the \textit{cognitivism} and \textit{constructivism} theories of learning \citep{jones2015networked,harasim2017learning}. For example, in a research on cognitive psychology, subjects were shown the sentence, \textit{The window is not closed}. Later, most of them recalled the sentence as \textit{The window is open} \citep{cross1998learning}, indicating that people build a mental image or a cognitive map, a \textit{schema}, in the learning process, and later build new knowledge in connection to what they already know \citep{mayes2013technology,harasim2017learning}.

Peer interaction is established as an important way people learn~\citep{topping2017effective,johnson2009educational}. In the process of collaboration, learning occurs as individuals build and improve their mental models through discussion and information sharing with their peers in small groups~\citep{davidson2014boundary,leidner1995use}. Research on learning communities has investigated how social and academic networks contribute to students' learning experiences and outcomes~\citep{brouwer2017emergent,smith2015magnets,smith2009learning,celant2013analysis}. In particular, study-related, task-dependent academic support relationships have been shown to lead to significantly higher academic performance~\citep{gavsevic2013choose,frank2008social}. In soft-skill development, \cite{fidalgo2015using} explored peer-interaction effects in developing teamwork competency in small groups, and found a direct relation between quantities of interaction and individual performance.

In this paper, we explore the scope of peer-induced \textit{speaking skill} development, namely, the influence of peer interactions in small groups towards improving one's speaking competencies. The ability to communicate ideas and thoughts effectively is valued to be an important skill throughout professional and casual interaction settings. Given this, can individuals improve their speaking skills by interacting with peers in small groups? If so, can such effects be extracted from data objectively, and then correctly modeled? To the best of our knowledge, there have been limited studies conducted in the past in this particular context. Part of the problem is in identifying or indeed generating relevant datasets. In particular, the monitoring and evaluation of peer effects objectively require data with quantifiable evidence. However, keeping a structured history of peer interactions had proved difficult in traditional offline studies, leading to the use of questionnaires and self-reported data \citep{lomi2011some,huitsing2012must}. 

With the recent advent of online communication, it is now commonplace to track detailed specifics of the participants' interactions from their online footprint, such as from Massive Open Online Courses' (MOOC) discussion forums~\citep{tawfik2017nature,brinton2014learning}. Online learning communities and other computer supported collaborative learning platforms thus facilitate objective studies of learner interaction and associated outcomes~\citep{jones2015networked,o2012computer,dado2017review,russo2005prestige, palonen2013patterns}. Consequently, our first contribution in this paper is in constructing a novel dataset of six online communities, where the participants focus on developing their \textit{speaking} skills. The 158 participants in the six communities record video speeches in response to common job interview prompts, and subsequently exchange feedback comments and performance ratings with peers in their respective communities. They respond to 5 prompts across 10 days, thereby generating a unique \textit{temporal} dataset of comment-based interactions and speaking performance ratings. In the dataset, the participants leave a comment only after they have watched a video of a peer. In such a pairwise interaction, Participant A can potentially learn from the feedback given by Participant B, and Participant B can also learn from Participant A's public speaking approach, good practices, and mistakes by watching the video. Moreover, the meta-cognitive task of explaining something through feedback can also clarify Participant B's own understanding~\citep{de2015exploring,roscoe2008tutor}. In line with the collaborative model of peer-induced learning, one might expect these interactions to help the participants gradually build mental models of the good practices of speaking~\citep{davidson2014boundary}. Knowledge regarding speaking skills has a tacit nature, which in turn can be developed through observation, imitation, and practice~\citep{busch2008tacit,chugh2013workplace}. Furthermore, knowledge is a non-rival good and can be traded without decreasing the level possessed by each trader. Therefore, such mutually beneficial interactions can be anticipated to lead to a gradual accumulation of speaking skills in the communities --- as measured by the performance ratings --- eventually driving the communities towards skill homogeneity as time goes on. Whether or not such gradual buildup of skills happens in a small group setting has not been reported in learning community literature, and that is a gap we address in this study. The temporal nature of our dataset facilitates such exploration.

Interactions in a learning platform are well suited to be modeled as a complex network, where the learners are denoted as nodes and their interactions as edges. This naturally leads to the use of Social Network Analysis (SNA) tools to gain insights on how various network parameters affect networked learning. For example, network position parameters (closeness and degree centrality, prestige etc.) have been shown to influence final grades~\citep{putnik2016analysing,cho2007social} and cognitive learning outcomes~\citep{russo2005prestige} of the learners. However, SNA is concerned with the \textit{structure of relational ties} between a group of actors~\citep{carolan2013social}, and not with any intrinsic measures of `performance' or `merits' of those actors. For example, if a person has a lot of followers on Twitter, the in-degree measure will be high, and one would say the person has a high in-degree centrality (in lay terms, enjoys a prominent position in the network) \citep{newman2010networks}. This intuition of `importance' is extracted from the connection topology, not from any intrinsic measure of how good the person's tweets actually are. Therefore, in a learning setting, researchers usually apply correlation analysis and other statistical methods to investigate the relations between SNA measures and performance/learning outcomes~\citep{dado2017review}. Recent developments in the emerging field of Graph Signal Processing (GSP) provide a novel way to integrate such `performance' measures (signals) with network graphs~\citep{shuman2013emerging}. The idea is to attach a signal value to every node in the graph, and process the signals atop the underlying graph structure. For example, \cite{huang2016graph} used GSP to study brain imaging data, where the brain structure is modeled as a graph and the brain activities as signals. \cite{deri_nyc_taxi} exploited GSP to study vehicle trajectories over road networks. Other recent application domains include Neuroscience \citep{rui2017simultaneous}, imaging \citep{pang2016graph,thanou2016graph}, medical imaging \citep{kotzagiannidis2016graph}, to name a few. In a learning setting, these developments in GSP open up the opportunity to model learners' academic grades or similar outcome measures as \textit{graph signals} on top of a peer interaction network. Consequently, our second contribution in this paper is in mining evidences of peer-influence and buildup of speaking skills using novel Graph Signal Processing based techniques. More formally, we ask (1) whether the learners' ratings are affected by their interaction patterns and the ratings of the peers they interact with, and (2) whether there is any gradual buildup of speaking skills in the communities towards homogeneity.

Towards answering the aforementioned research questions, we take two approaches. \textit{First}, we measure the smoothness of the graph signals (speaking ratings) as the participants interact temporally across 5 prompts. We show that with time, people's ratings come closer to the ratings of the peers they interact with and the ratings also increase on average. This suggests that the communities in the dataset gradually approach homogeneity in terms of the performance ratings of the interacting participants. \textit{Second}, we approach the mining of peer-influence as a prediction formulation. Let us introduce the idea through a stylized scenario. In a classroom setting, a teacher can track the test scores of a student and make a prediction of his/her future score by running a linear regression through previous scores, providing a baseline trend. Instead, if the teacher also takes into account the scores of the student's peers (those that the student interacts with), we demonstrate that the prediction of future scores for both the student and the peers outperform the baseline. More formally, a network-consensus prediction model outperforms a network-agnostic model, a trend we demonstrate to be true in all six communities in our generated data. This corroborates previous findings that interaction quantities and peers' grades/performance outcomes impact learners' own performance outcomes~\citep{fidalgo2015using, hoxby2000peer}, and we show that they hold for speaking skill development as well.

We further proceed to model and simulate the peer-effects observed in the dataset. From an economic lens, various models for knowledge, information or opinion flow across a network has been explored in previous literature~\citep{golub2017learning,cowan2004network,gale2003bayesian}. We introduce a model of peer-induced knowledge propagation (in terms of speaking performance ratings) based on consensus protocols for network synchronization and distributed decision making~\citep{Olfati_Tutorial_2007}. ~\cite{olfati2004consensus} studied such consensus algorithms and presented relevant convergence analysis. We modify their protocol to suit the characteristics of the communities of our dataset. We discuss how the proposed diffusion dynamics capture the peer-influence and the gradual buildup of performance ratings towards community-wide homogeneity, as observed in our dataset. 

In this context, our contributions can be summarized as follows:
\begin{itemize}
\item Constructing a dataset of six online communities that allow studying longitudinal peer-effects in speaking skill development;
\item Yielding new evidences for skills gradually building up in learning communities towards homogeneity, corroborating previous findings of positive impacts of peer-influence, and modeling the observations via diffusion dynamics;
\item Employing novel Graph Signal Processing tools for extracting the insights.
\end{itemize}

\section{Dataset}
\subsection{Data Collection}
The data comes from a ubiquitous online system, ROC Speak [reference omitted], that gives people semi-automated feedback on public speaking. 158 participants, aged 18 to 54 years, were hired from Amazon Mechanical Turk. All of them were native English speakers, and came from a variety of professional and educational backgrounds. The participants were randomly assigned into 6 groups labeled 1 through 6, which had 26, 31, 26, 30, 22 and 23 participants respectively. They were assigned a common goal of improving their speaking skills, towards which they were given 5 common job interview prompts in 10 days. Responses to these prompts were recorded via webcam as videos (typically 2 minutes in length) and stored in the platform. The study was conducted upon approval from the authors' University IRB. Note, that in what is to follow, we use the terms groups and communities interchangeably.

\subsubsection{Automated Feedback} For groups 2, 4, 6, the system generated automatic feedback on smile intensity, body movement/gesture, loudness, pitch, unique word count, word cloud and instances of weak language use, as shown in Figure \ref{fig:interface}. Groups 1, 3, 5 did not receive any automated feedback.

\subsubsection{Feedback Comments and Ratings}
In all six groups, the participants exchanged feedback with their peers in their respective groups. They were required to give feedback to at least three peers in each prompt, whom they could choose at will from a feed of all their peers' videos. Each feedback comprised of (1) at least three comments and (2) performance ratings on a 1-5 Likert scale. 

The participants were not given any explicit instruction on what aspects of speaking skills to give comments on. However, the user interface allowed them to tag the comments whether they were on friendliness, volume modulation or use of gestures. The comments were mostly focused on speech delivery. Here, we do not concern ourselves with an analysis of the specific contents of the comments, instead refer to each comment as an `interaction' unit. 

\begin{figure}[t]
        \centering    
            {\includegraphics[width=0.7\linewidth]{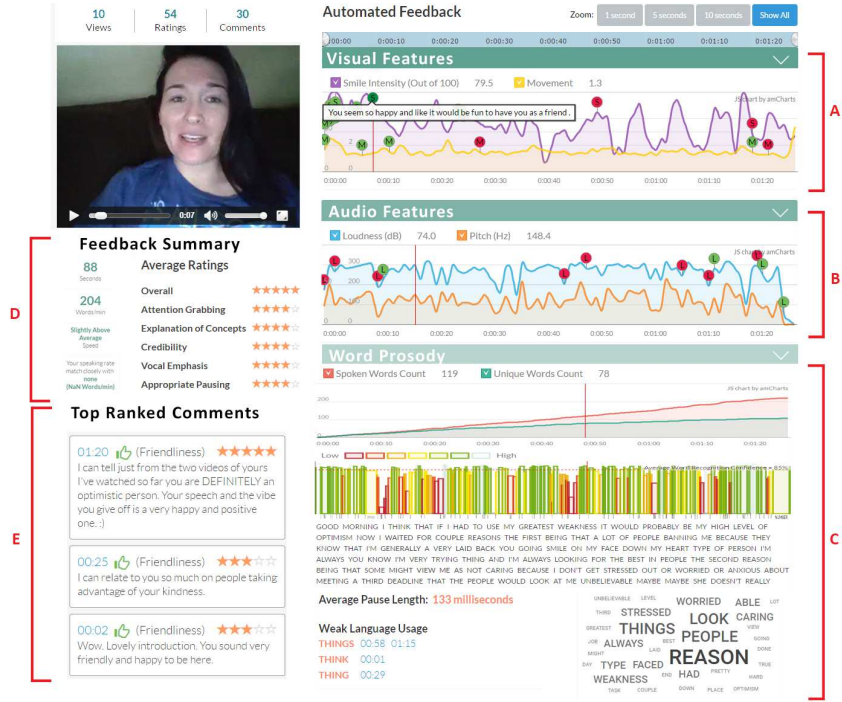}}
                              
        \caption{A snapshot of the ROC Speak feedback interface. The page shows automatically generated measurements for (A) smile intensity, gestural movements, (B) loudness, pitch, (C) unique word count, word recognition confidence, a transcription of the speech, word cloud, and instances of weak language. Additionally, the peer-generated feedback are shown in (D) a Feedback Summary section, and (E) a top ranked comments section classified by usefulness and sentiment. In our study, we use the overall ratings given by the peers, as shown in segment (D).}
        \label{fig:interface}
    \end{figure}

When uploading a video, participants could select 5 qualities from a list of 23 qualities on which they wanted performance ratings from their peers. For example, the video in Figure \ref{fig:interface} asks for ratings on the qualities of Attention Grabbing, Explanation of Concepts, Credibility, Vocal Emphasis, Appropriate Pausing, as shown in segment D of the figure. In addition to these customized rating categories, all videos received `overall' delivery performance ratings from the peers, as shown in segment D of Figure \ref{fig:interface}. The participants were at liberty to use their judgments in giving ratings to their peers. For our analysis, we use the average `overall' rating each video received from the peers and refer to this average value as `peer rating', `performance rating'  or `speaking rating' interchangeably in the sequel, considering it to be an objective abstraction of the participant's overall performance or skill level in delivering the speech.

\subsection{Dataset Summary}
All of the 158 participants in the 6 groups completed the study by recording at least one video to all 5 prompts --- generating a total of 817 videos. The 6 groups have 25,665 peer-generated ratings in total. Out of them, 5,053 are `overall' performance ratings that we use in this study. Therefore, each video received `overall' ratings from 6.19 peers on average.  The dataset has 14,285 comments in total, with an average of 17.49 comments per video.

\section{Methodology}
In this section, we detail the methods of mining peer-influence evidence in the dataset. As argued in the Introduction, we are interested to see if the interactions impact the development of speaking-related skills in the six groups. To reiterate, we examine (1) whether the learners' ratings are affected by their interaction patterns and the ratings of the peers they interact with, and (2) whether there is any gradual buildup of speaking skills in the communities towards homogeneity.

	\subsection{Performance Ratings as Graph Signals}
\begin{figure}[h]
        \centering    
            {\includegraphics[width=0.7\linewidth]{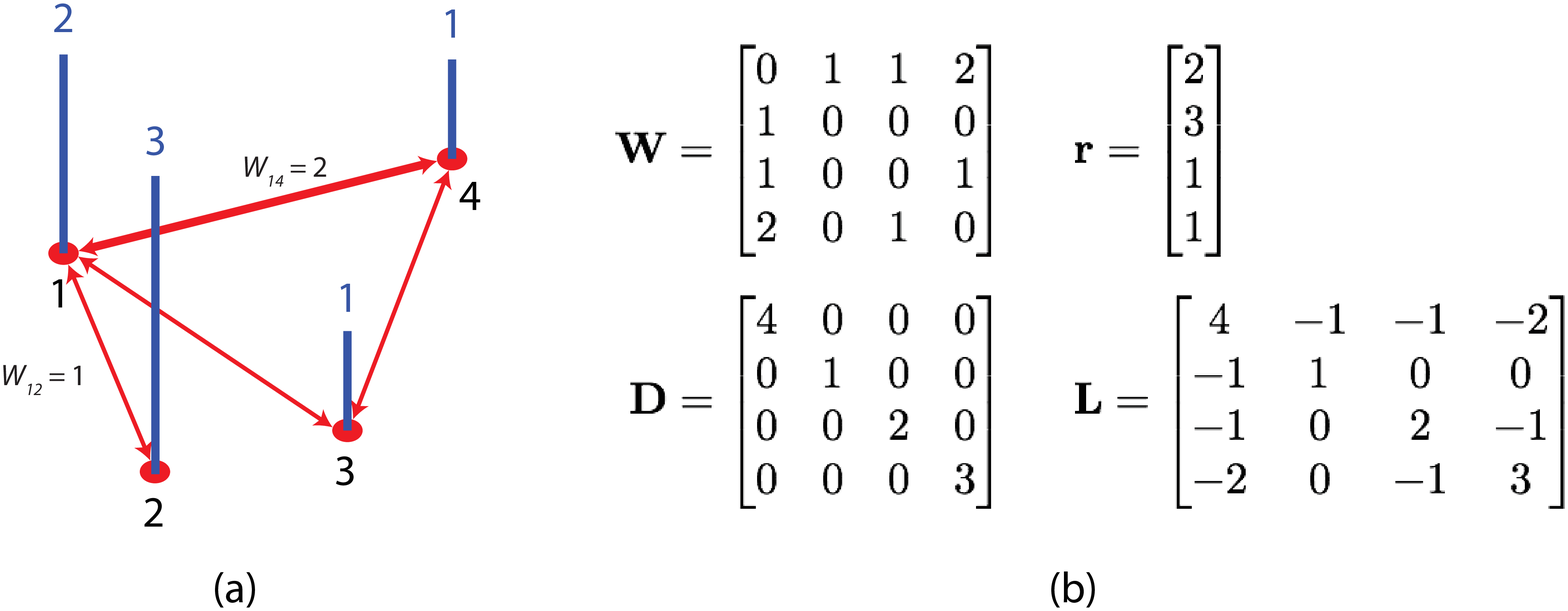}} 
                              
        \caption{(a) A network graph representation of an example community with $N=4$ participants depicted as red nodes. The edges denote interactions in the form of feedback comments, with the edge weights corresponding to the number of interactions between nodes. As indicated by edge weights, participants $1$ and $4$ interact 2 times, while all other nonzero pairwise interactions take place only once. The blue bars on top of the nodes represent average speaking ratings as \textit{graph signals}, and can take any value between 1 and 5.  (b) The Adjacency matrix $\mathbf{W}$ captures the number of interactions between participants $i$ and $j$ in its $ij$\textsuperscript{th} element. Graph signal $\mathbf{r}$ captures everyone's speaking performance ratings. The diagonal degree matrix $\mathbf{D}$ has node $i$'s number of interactions in its $ii$\textsuperscript{th} entry. The Laplacian $\mathbf{L}$ is computed by $\mathbf{D}-\mathbf{W}$. In our dataset, $\mathbf{W}$, $\mathbf{r}$, $\mathbf{D}$ and $\mathbf{L}$ vary across prompts $p$, and also across the six groups. }
        \label{fig:basic_model}
    \end{figure}

        Consider modeling an online community's participants and their interactions through a network graph. To that end, we naturally identify each participant with a node in the said graph. As discussed in the Introduction, each comment-based interaction can potentially influence both the commenter and the receiver. The receiver can benefit from the comment itself. On the other hand, the commenter can learn from watching the peer's video and from the meta-cognitive task of providing a feedback. Therefore, each interaction is represented via an undirected edge, acknowledging mutual benefit for the two nodes. Exchanges of multiple comments are encoded through integer-valued edge weights; more formally, consider an undirected, weighted graph $G(\mathcal{N},\mathcal{E},\mathbf{W}^{(p)})$ representing this network at $p$\textsuperscript{th} prompt with a node set $\mathcal{N}$ of known cardinality $N$ (i.e., the total number of participants in the group---each of the groups in the dataset have their own cardinality), and the edge set $\mathcal{E}$ of unordered pairs of elements in $\mathcal{N}$. The so-called symmetric weighted adjacency matrix is denoted by $\mathbf{W}^{(p)} \in \mathbb{R}^{N \times N}$, whose $ij$\textsuperscript{th} element represents the total number of comments that participants $i$ and $j$ have given to each other in the $p$\textsuperscript{th} prompt. Figure \ref{fig:basic_model} illustrates the construction of $\mathbf{W}$ using a toy example. Since the interaction patterns tend to change from prompt to prompt, so does the connectivity pattern (i.e., the topology) of the resulting graph and hence the explicit dependency of the weights in $\mathbf{W}^{(p)}$ with respect to $p$.

Next, we incorporate the participants' rating information in the form of graph signals. Here, a \textit{graph signal}~\citep{shuman2013emerging} is a vertex-valued network process that can be represented as a vector of size $N$ supported on the nodes of $G$, where its $i$\textsuperscript{th} component is the rating of node $i$. As explained in the Dataset section, the participants are given overall performance ratings by their peers, and we take a participant's average overall rating in each prompt as the graph signal value. Thus, we collect the ratings of the $p$\textsuperscript{th} prompt in a vector $\mathbf{r}^{(p)} \in \mathbb{R}^{N}$, where $r_{i}^{(p)}$ is $i$\textsuperscript{th} participant's rating. This representation is visualized in Figure \ref{fig:basic_model}. Under the natural assumption that signal properties are related to graph topology, the goal of Graph Signal Processing is to develop models and algorithms that fruitfully leverage this relational structure, making inferences about these signal values when they are only partially observed~\citep{shuman2013emerging}.

		\subsection{Total Variation Analysis}
We explore whether the communities gradually approach homogeneity in performance ratings, by conducting a total variation analysis. Total variation is a measure of smoothness of the graph signals (ratings), the idea being, if non-rival interactions allow participants to gradually achieve ratings closer to their peers, then ratings will become smoother or more homogeneous across the network. 

Given the network adjacency matrix $\mathbf{W}^{(p)} \in \mathbb{R}^{N \times N}$, the degree (number of links) of participant $i$ at the $p$\textsuperscript{th} prompt is the total number of comments exchanged in that prompt, defined as $d^{(p)}_{i} := \sum_{j} W^{(p)}_{ij}$. Using these definitions, we construct the Laplacian matrix for any prompt $p$ according to $\mathbf{L}^{(p)} := \mathbf{D}^{(p)}-\mathbf{W}^{(p)}$, where $\mathbf{D}^{(p)}:=\text{diag}(\mathbf{W}^{(p)}\mathbf{1}_{N})$ is a diagonal matrix with elements corresponding to the nodes' degree, and $\mathbf{1}_{N}$ are the vector of ones of length $N$~\citep{graph_spec}. Figure \ref{fig:basic_model} illustrates the construction of the degree and Laplacian matrices. For a graph signal $\mathbf{r}$, one can utilize the graph Laplacian to compute the total variation (TV) of the participants' ratings thus,

\begin{equation} \label{total_variation} %centering
\text{TV} (\mathbf{r}) := \mathbf{r}^{T}\mathbf{L}\mathbf{r}=\sum_{i,j=1,i>j}^{N}W_{ij}(r_{i}-r_{j})^{2},
\end{equation}
where $r_{i}$ and $r_{j}$ reflect the ratings of participants $i$ and $j$, while weights $W_{ij}$ account for their volume of interaction. The total variation in Eq.~\eqref{total_variation} is often referred to as a smoothness measure of the signal $\mathbf{r}$ with respect to the graph $G$. If two nodes $i$ and $j$ do not interact in a prompt, $W_{ij}$ is $0$ and therefore $(r_{i}-r_{j})^{2}$ makes no contribution to the sum in Eq.~\eqref{total_variation}, while increasing interaction leads to more contributions from the term $(r_{i}-r_{j})^{2}$. If $\mathbf{r}$ differs significantly between pairs of nodes that show strong patterns of interaction, then $\text{TV} (\mathbf{r})$ is expected to be large. On the other hand, if ratings vary negligibly across connected nodes, then the graph signal is smooth and $\text{TV} (\mathbf{r})$ takes on small values. Increasing convergence of ratings across interacting participants leads to a decreasing total variation. However, TV does not capture the overall trend of the ratings---whether the participants' ratings are improving or not. Therefore, to complement the total variation analysis, we calculate the network-wide average ratings at each prompt as

\begin{equation} \label{average}
\bar{r}^{(p)} :=\frac{1}{N}\sum_{i=1}^N r_i^{(p)} ,
\end{equation}
where $r_i^{(p)}$ is the rating of the $i$\textsuperscript{th} participant in the $p$\textsuperscript{th} prompt.

		\subsection{Prediction Under Smoothness Prior}
       
In the Introduction, we presented a scenario of a teacher tracking the performance rating evolution of students to illustrate our approach of prediction formulation towards shedding light on peer-effects. More formally, we formulate a convex optimization problem to predict future performance ratings. We compare prediction errors between two models: (i) an ordinary least-squares regression model using only past performance ratings of the participant, i.e., a model which is agnostic to the network effects on the participant; and (ii) augmenting the least-squares predictor with a smoothness regularization term encouraging network-wide consensus. In other words, the second model considers the ratings of one's peers and the quantities of interaction along with one's own records. It is important to note that we do not seek to fit the data to find the best possible regression model for predicting future ratings, rather, we intend to compare prediction errors between network-agnostic and consensus-based regression models to illustrate our point that the network has an impact on individual outcomes.

We use the first four prompts in our data for training purposes, and predict the 5\textsuperscript{th} prompt's ratings for all the participants in the six groups. Specifically, we collect the rating signal vectors $\mathbf{r}^{(p)} \in \mathbb{R}^{N}$ for prompts $p=1,2,3,4$ to form the training set and predict the ratings at prompt $p=5$ using two different linear regression models:
\begin{equation} \label{e:linear_predict}
\hat{\mathbf{r}}=\boldsymbol{\beta}_{0} + \boldsymbol{\beta}_{1} p \\
\end{equation}
and
\begin{equation} \label{e:nonlinear_predict}
\hat{\mathbf{r}}=\boldsymbol{\beta}_{0} + \boldsymbol{\beta}_{1} p + \boldsymbol{\beta}_{2} \sqrt[]{p},
\end{equation}
where $\hat{\mathbf{r}} \in \mathbb{R}^{N}$ is the estimated ratings at $p$\textsuperscript{th} (in our case, $5$\textsuperscript{th}) prompt. For \eqref{e:linear_predict} and \eqref{e:nonlinear_predict}, the parameters $\boldsymbol{\beta}_{0}, \boldsymbol{\beta}_{1}, \boldsymbol{\beta}_{2}  \in \mathbb{R}^{N}$ are learned through the following regularized least-squares criteria respectively:

\begin{equation} \label{e:opt:smooth}
\{\hat{\boldsymbol{\beta}}_{0}, \hat{\boldsymbol{\beta}}_{1}\} = 
\underset{\boldsymbol{\beta}_{0},\boldsymbol{\beta}_{1}}{\operatorname{argmin} } \frac{1}{2m} \sum_{p=1}^{m} \left \lVert \mathbf{r}^{(p)}-(\boldsymbol{\beta}_{0} + \boldsymbol{\beta}_{1} p) \right \Vert^{2}
+ \lambda \sum_{p=1}^{m} (\boldsymbol{\beta}_{0} + \boldsymbol{\beta}_{1} p)^{T} \mathbf{L}^{(p)} (\boldsymbol{\beta}_{0} + \boldsymbol{\beta}_{1} p)
\end{equation}

and
\begin{align} \label{e:opt:smooth_non_linear}
\{\hat{\boldsymbol{\beta}}_{0}, \hat{\boldsymbol{\beta}}_{1}, \hat{\boldsymbol{\beta}}_{2}\} = 
\underset{\boldsymbol{\beta}_{0},\boldsymbol{\beta}_{1}, \boldsymbol{\beta}_{2}}{\operatorname{argmin} } \frac{1}{2m} &\sum_{p=1}^{m} \left \lVert \mathbf{r}^{(p)}-(\boldsymbol{\beta}_{0} + \boldsymbol{\beta}_{1} p + \boldsymbol{\beta}_{2} \sqrt[]{p}) \right \Vert^{2} \nonumber\\
&+ \lambda \sum_{p=1}^{m} (\boldsymbol{\beta}_{0} + \boldsymbol{\beta}_{1} p + \boldsymbol{\beta}_{2} \sqrt[]{p})^{T} \mathbf{L}^{(p)} (\boldsymbol{\beta}_{0} + \boldsymbol{\beta}_{1} p + \boldsymbol{\beta}_{2} \sqrt[]{p}) \nonumber
\\
&+ \mu (\left \lVert \boldsymbol{\beta}_{1} \right \Vert^{2} + \left \lVert \boldsymbol{\beta}_{2} \right \Vert^{2}).
\end{align}

Here, the regression model has the non-linear $\sqrt[]{p}$ term to capture any saturation effect at limiting time. We have $m=4$ since the training is done on the first four prompts' data. Let us break down the objective functions \eqref{e:opt:smooth} and \eqref{e:opt:smooth_non_linear} for clarity. The first summands in \eqref{e:opt:smooth} and \eqref{e:opt:smooth_non_linear} are data fidelity terms, which take into account each participant's individual trajectory of learning across the first $m$ prompts, and fits them to the respective linear regression models. Accordingly, the first terms can be attributed to a person's own talent or pace of learning, agnostic to the network. The second summands in \eqref{e:opt:smooth} and \eqref{e:opt:smooth_non_linear} incorporate the effects of the neighbors' ratings and the amount of interaction into a participant's learning curve. Notice that these terms are nothing else than smoothing regularizers of the form $\sum_{p=1}^{m} \text{TV}(\hat{\mathbf{r}})$, hence encouraging participant ratings prediction with small total variation. The tuning parameter $\lambda>0$ balances the trade-off between faithfulness to the past ratings data and the smoothness (in a total variation sense) of the predicted graph signals $\hat{\mathbf{r}}=\hat{\boldsymbol{\beta}}_{0} + \hat{\boldsymbol{\beta}}_{1}p$ and $\hat{\mathbf{r}}=\hat{\boldsymbol{\beta}}_{0} + \hat{\boldsymbol{\beta}}_{1} p + \hat{\boldsymbol{\beta}}_{2} \sqrt[]{p}$, and can be chosen via model selection techniques such as cross validation~\citep{friedman2001elements}. Parameter $\mu$ which is chosen along with $\lambda$ via leave-one-out cross validation, prevents overfitting via a shrinkage mechanism which ensures that none of $\boldsymbol{\beta}_{1}$ and $\boldsymbol{\beta}_{2}$ get to dominate the objective functions disproportionately~\citep{friedman2001elements}.
Equations \eqref{e:opt:smooth} and \eqref{e:opt:smooth_non_linear} are both convex, specifically unconstrained quadratic programs that can be solved efficiently~\citep{boyd2004convex} via off-the-shelf software packages. Here we use CVX \citep{grant2008cvx}, a MATLAB-based modeling system for solving convex optimization problems. An outline of the procedure is presented in Algorithm \ref{alg:make_traj}.

\begin{algorithm}[t]
\begin{flushleft}
	\textbf{Input:} Rating vectors $\{\mathbf{r}^{(p)}\}_{p=1}^{m}$ and Laplacian matrices $\{\mathbf{L}^{(p)}\}_{p=1}^{m}$\\

	\textbf{Output}: Predicted ratings at $p=(m+1)^{\text{th}}$ prompt
 \begin{algorithmic}
 	\State Choose parameters $\lambda$ and $\mu$ using cross validation over first $m$ prompts by solving \eqref{e:opt:smooth} (or \eqref{e:opt:smooth_non_linear}) using CVX
  \State Obtain parameters $\boldsymbol{\beta}_{0}, \boldsymbol{\beta}_{1}, \boldsymbol{\beta}_{2}$ by solving \eqref{e:opt:smooth} (or \eqref{e:opt:smooth_non_linear}) using CVX
  \State Predict the ratings at prompt $p=m+1$ using \eqref{e:linear_predict} (or \eqref{e:nonlinear_predict})
 \end{algorithmic}
\end{flushleft}

\caption{Prediction Under Smoothness Prior}\label{alg:make_traj}
\end{algorithm}

Setting $\lambda = 0$ in Eqns.~\eqref{e:opt:smooth} and~\eqref{e:opt:smooth_non_linear} reduces them to ordinary network-agnostic least-squares regression models. In simple terms, a participant's $5$\textsuperscript{th} prompt rating is then predicted by running linear regression over previous ratings, which is used as a baseline. For $\lambda \neq 0$ chosen by cross-validation, the network-effects are switched on in~\eqref{e:opt:smooth} and~\eqref{e:opt:smooth_non_linear}, and the optimization procedure determines parameters that minimize the total variation of the participant ratings in combination with least-squares regression. Peer effects are then determined by comparing the model with $\lambda \neq 0$ to that with $\lambda = 0$. Relative prediction errors defined as $\| \hat{\boldsymbol{\beta}}_{0} + 5 \hat{\boldsymbol{\beta}_{1}} - \mathbf{r}^{(5)} \|/\| \mathbf{r}^{(5)} \|$ and $\| \hat{\boldsymbol{\beta}}_{0} + 5 \hat{\boldsymbol{\beta}_{1}} + \sqrt[]{5} \hat{\boldsymbol{\beta}_{2}} - \mathbf{r}^{(5)} \|/\| \mathbf{r}^{(5)} \|$ respectively for \eqref{e:opt:smooth} and \eqref{e:opt:smooth_non_linear} are reported in Table \ref{tab:pred_error}.

\subsection{Modeling and Simulation}
We proceed to develop a model for the participant ratings' evolution which facilitates simulation of the network process. We model the temporal update of the ratings $\mathbf{r}^{(p)}$ as a diffusion process on the graph $G$, where the trade of feedback can take place without decreasing the individual level of knowledge/speaking ratings. We impose a positive drift to the ratings in order to model the fact that a learner can accumulate an understanding of skills (i.e., build mental models of better practices in speaking) from various sources such as automated feedback, peer feedback, experience of watching peers' videos, and other external resources. In a real network, people will have different learning rates, their individual talents will differ, the external inflow of information will not be consistent, and in some less probable cases, they may forget information as well. To lump all these noisy yet positively skewed effects into one variable, we model the positive drift $\boldsymbol{\epsilon}$ as a Gaussian random variable with a positive mean $\mu>0$. Superposition of random effects is well modeled by a Gaussian random variable by virtue of the Central Limit Theorem. The ratings received by participants lie between 1-5, so we introduce an explicit control procedure to bound the ratings in our model. Thus the evolution of ratings can be modeled via the Laplacian dynamics
\begin{equation}
\label{e:model_online}
\mathbf{r}^{(p+1)} =  \mathcal{P}_{[r_{\min},r_{\max}]} \left( \mathbf{r}^{(p)} - c \mathbf{L}^{(p)} \mathbf{r}^{(p)} + \boldsymbol{\epsilon} \right),
\end{equation}
where $\mathbf{r}^{(p)}$ is the ratings vector of $p$\textsuperscript{th} prompt, $c \in (0,1/d_{max})$ is the diffusion constant and $d_{max}$ is the maximum degree of nodes at the corresponding prompt, $\mathbf{L}^{(p)}$ is the Laplacian matrix of the graph $G$ at the $p$\textsuperscript{th} prompt, $\boldsymbol{\epsilon}$ is a Gaussian random vector with mean $\mu>0$ and given variance $\sigma^2$, and $\mathcal{P}_{[r_{\min},r_{\max}]}(\cdot)$ is a projection operator onto the interval $[r_{\min},r_{\max}]$. For ROC Speak one has $r_{\min}=1$ and $r_{\max}=5$.

\begin{figure}
        \centering    
            {\includegraphics[width=0.5\linewidth]{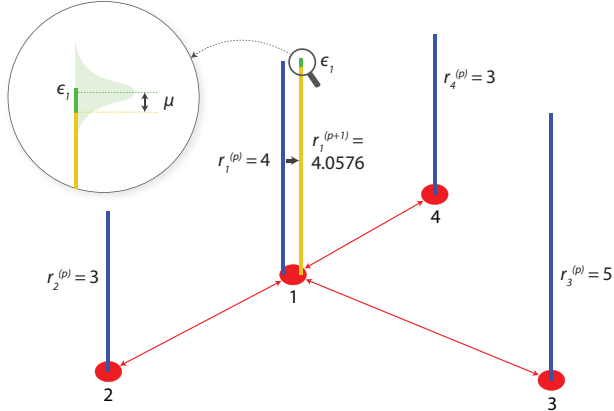}} 
                              
        \caption{An example demonstrating how participant $1$'s rating in the $(p+1)$\textsuperscript{th} prompt, $r^{(p+1)}_{1}$, is calculated from the participant's own rating in the $p$\textsuperscript{th} prompt, $r^{(p)}_{1} = 4$, single interactions with participants $2$, $3$ and $4$, and a positive drift $\epsilon_{1}$. The $p$\textsuperscript{th} prompt ratings for the participants are shown by blue bars. Here, participants $2$ and $4$ both have ratings of $3$ in the $p$\textsuperscript{th} prompt, and participant $3$ has a rating of $5$. The net diffusion effect with peers takes participant $1$'s rating from 4 down to $3.99$ (denoted by a yellow bar). However, a Gaussian random positive drift $\epsilon_{1} = 0.0676$ (denoted by a green bar on top of the yellow bar) represents a sample realization of participant $1$'s gathering of understanding of speaking skills from sources external to peer feedback, and pushes his rating in the $(p+1)$\textsuperscript{th} prompt to $r^{(p+1)}_{1} = 3.99 + 0.0676 = 4.0576$. The magnified plot reflects the probability distribution of participant $1$'s rating at $(p+1)$\textsuperscript{th} prompt which is centered around $3.99 + \mu$, with mean $\mu$. This also shows that the rating in $(p+1)$\textsuperscript{th} prompt can take any value in the interval $[r_{\min},r_{\max}]$, but with a higher probability it is closer to $3.99 + \mu$.}
        \label{fig:diffusion_example}
    \end{figure}

Figure \ref{fig:diffusion_example} demonstrates the overall idea pictorially. To understand the chosen dynamics, disregard the projection operator in \eqref{e:model_online} for the sake of a simpler argument. Then notice that the update  $\mathbf{r}^{(p+1)}=\mathbf{r}^{(p)} - c \mathbf{L}^{(p)} \mathbf{r}^{(p)}$ represents a Laplacian-based network diffusion process~\citep{olfati2004consensus}, where the future rating of a given participant depends on the ratings of his/her peers in the current prompt ($\mathbf{r}^{(p)}$) and the nature of the interactions taking place in the learning community ($\mathbf{L}^{(p)}$). Focusing on the $i$\textsuperscript{th} participant recursion in \eqref{e:model_online} (modulo the projection operator), one obtains the scalar update
\begin{equation}
\label{e:model_online_user_i}
r^{(p+1)}_{i} =   (1 - c d_{i})r^{(p)}_{i} + c \sum_{j\in \mathcal{V}} W^{(p)}_{ij} r^{(p)}_{j} + \epsilon_{i},
\end{equation}
where $r^{(p)}_{i}$ is $i$\textsuperscript{th} participant rating at the $p$\textsuperscript{th} prompt. In obtaining \eqref{e:model_online_user_i}, we have used the definition of graph Laplacian, i.e., $\mathbf{L}^{(p)}:=\mathbf{D}^{(p)}-\mathbf{W}^{(p)}$. The quantity $(1 - c d_{i}) r^{(p)}_{i} + c \sum_{j} W^{(p)}_{ij} r^{(p)}_{j}$ is a weighted average of participant $i$ and his/her neighbors' ratings, with each neighbor's weights being proportional to the number of interactions that s/he has with participant $i$. This way, the model captures peer-influenced buildup of ratings across the network where the diffusion constant $c$ is a relatively small number. Further intuition can be gained by interpreting $\mathbf{r}^{(p+1)}=\mathbf{r}^{(p)} - c \mathbf{L}^{(p)} \mathbf{r}^{(p)}$ as a gradient-descent iteration to minimize the total variation functional $\text{TV}(\mathbf{r}):=\mathbf{r}^T\mathbf{L}^{(p)}\mathbf{r}$ in \eqref{total_variation}. This suggests that \eqref{e:model_online} will drive the ratings towards a consensus of minimum total variation. 

We test the proposed model in Eq.~\eqref{e:model_online} by running a numerical simulation that synthetically generates participant ratings across prompts. We synthetically generate graphs with structural properties resembling our dataset. Given that in our platform, participants can, in principle, send communication at random, we generate Erd\H{o}s-R\'enyi random graphs for each prompt with the average number of nodes in the communities ($N = 26$ nodes) \citep{erdos1960evolution}. Each edge is included in the graph with probability $p = 0.5$, so that the expected number of edges $\binom{N}{2}$$p$ matches the number of edges in the dataset. We set the positive drift term $\boldsymbol{\epsilon}$to have a mean $\mu=0.05$ and standard deviation $\sigma=0.1$ based on the ROC Speak average ratings behavior, and select $c = 0.01$.

\section{Results}
		\subsection{Convergence of community-wide performance with increasing interactions}
        
\begin{figure}[h]
        \centering    
%             {\includegraphics[width=1\linewidth]{total_var_and_ratings.eps}} 
{\includegraphics[width=1\linewidth]{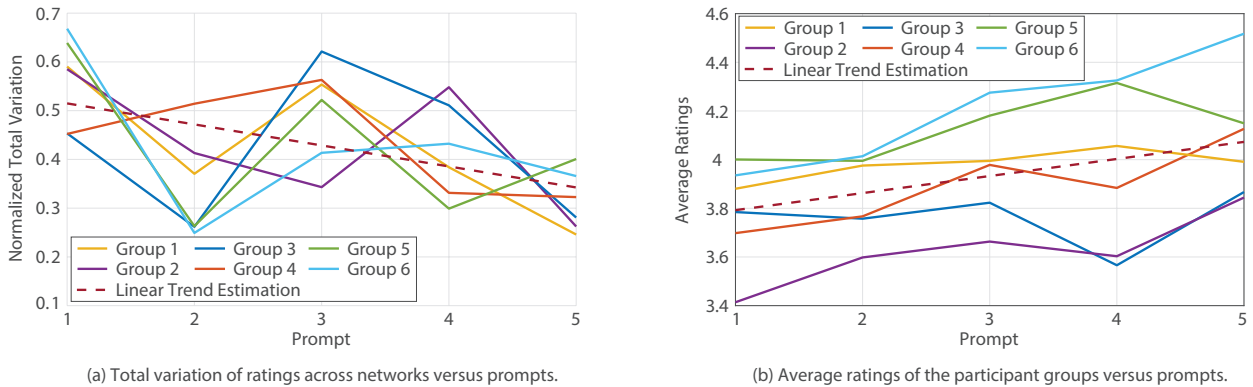}} 
                              
        \caption{Total variations and average ratings as functions of prompts. In (a), the total variation decreases across prompts while (b) shows the network-wide average ratings increase. The dashed lines are the linear trend estimations across prompts.}
        \label{fig:tot_var_over_rat}
    \end{figure}

        We evaluate the total variation of the rating signals $\mathbf{r}^{(p)}$ across the ROC Speak network at the end of each prompt [cf.~\eqref{total_variation}]. We collect $i$\textsuperscript{th} community's TV of the 5 prompts in a vector $\textbf{TV}_{i} = (\text{TV}_{i}^{(1)},\text{TV}_{i}^{(2)}, \ldots ,\text{TV}_{i}^{(5)})^{T}$ and plot the normalized total variation as $\textbf{TV}_{i}/\lVert\textbf{TV}_{i}\rVert$ over prompts in Figure \ref{fig:tot_var_over_rat}(a), where $\lVert . \rVert$ maps a vector to its Euclidean norm. The normalized total variations mostly decrease from the third prompt onwards. The dashed plot in Figure~\ref{fig:tot_var_over_rat}(a) indicates the linear trend of the average across the six groups ($\frac{1}{6} \sum_{i=1}^{6} \textbf{TV}_{i}$). A diminishing trend is apparent, with a moderate yet significant slope of $-0.04$ ($p < 0.01$), suggesting that interacting participants gradually converge in terms of ratings.

%and \ref{fig:sim}(a)

       % \subsubsection*{\textbf{Overall Trajectory of Ratings}}
In addition, we plot the network-wide average ratings defined in Eq.~\eqref{average} as a function of prompts $p$ in Figure~\ref{fig:tot_var_over_rat}(b). It is evident that all six groups have a trend of improvement in average ratings across prompts. The dashed plot indicates the average linear trend estimation across all six groups, which has a moderate and significant slope of $0.07$ ($p < 0.05$). Taken together, these findings provide strong evidence that communities gradually approach homogeneity in terms of (improved) performance with increasing interactions among participants.

		\subsection{Improved prediction of ratings as a function of incorporating peer effects}
As explained in the Methods section, we compare the predictions against the original ground-truth ratings of the final prompt to report the prediction errors. 
The results are summarized in Table \ref{tab:pred_error}. The first and third columns of the table show the prediction errors obtained by our proposed consensus-based predictors in \eqref{e:opt:smooth} and \eqref{e:opt:smooth_non_linear}. The second and fourth columns refer to the baseline predictors of ordinary least-squares regression, which are computed by making the smoothness regularization term $\lambda=0$ in \eqref{e:opt:smooth} and \eqref{e:opt:smooth_non_linear}. Table \ref{tab:pred_error} shows that the consensus-based predictors ($\lambda \neq 0$) outperform their baseline counterparts ($\lambda = 0$) in all six communities. In particular, the consensus-based approach relatively improves the predictions by $23\%$ averaged over six groups. In other words, the ratings of one's peers and the associated interaction patterns help predict his/her performance ratings better. This, in turn, supports the idea that interactions with the neighbors indeed impact one's learning outcomes.

\begin{table}
\centering
\caption{Fifth prompt prediction errors comparison between consensus-based $(\lambda \neq 0)$ and network-agnostic $(\lambda = 0)$ frameworks, for regression models with (a) linear features using~\eqref{e:linear_predict} and (b) non-linear features using~\eqref{e:nonlinear_predict}. Bold errors denote better prediction results. The optimization functions for the regression models are elaborated through equations~\eqref{e:opt:smooth} and \eqref{e:opt:smooth_non_linear} in the Methods section.}
\begin{tabular}{|c||c|c||c|c|}
 \hline
 & \multicolumn{2}{|c||}{(a) Prediction errors with linear features} & \multicolumn{2}{|c|}{(b) Prediction errors with non-linear features}\\
 \hline

  & \begin{tabular}[x]{@{}c@{}}Consensus-based\\($\lambda \neq 0$)\end{tabular}  &  \begin{tabular}[x]{@{}c@{}}Network-agnostic\\($\lambda = 0$)\end{tabular} &  \begin{tabular}[x]{@{}c@{}}Consensus-based\\($\lambda \neq 0$)\end{tabular} &  \begin{tabular}[x]{@{}c@{}}Network-agnostic\\($\lambda = 0$)\end{tabular}\\
  \hline 
    \hline 
  Group 1 & \textbf{9.85\%} & 14.52\% & \textbf{7.83\%} & 8.18\%\\
  \hline 
  Group 2 & \textbf{10.93\%} & 18.67\% & \textbf{10.72\%} & 12.75\%\\
  \hline 
  Group 3 & \textbf{12.42\%} & 15.62\% & \textbf{9.95\%} & 13.14\% \\
  \hline 
  Group 4 & \textbf{11.26\%} & 15.4\% & \textbf{11.31\%} & 12.81\% \\
   \hline 
  Group 5 & \textbf{12.29\%} & 13.99\% & \textbf{10.02\%} & 12.57\% \\
  \hline 
  Group 6 & \textbf{7.04\%} & 11.43\% & \textbf{7.06\%} & 10.47\% \\
 \hline
\end{tabular}
\label{tab:pred_error}

\end{table}

\subsection{Capturing the dynamics of the interactions}
\begin{figure}[h]
        \centering    
            {\includegraphics[width=1\linewidth]{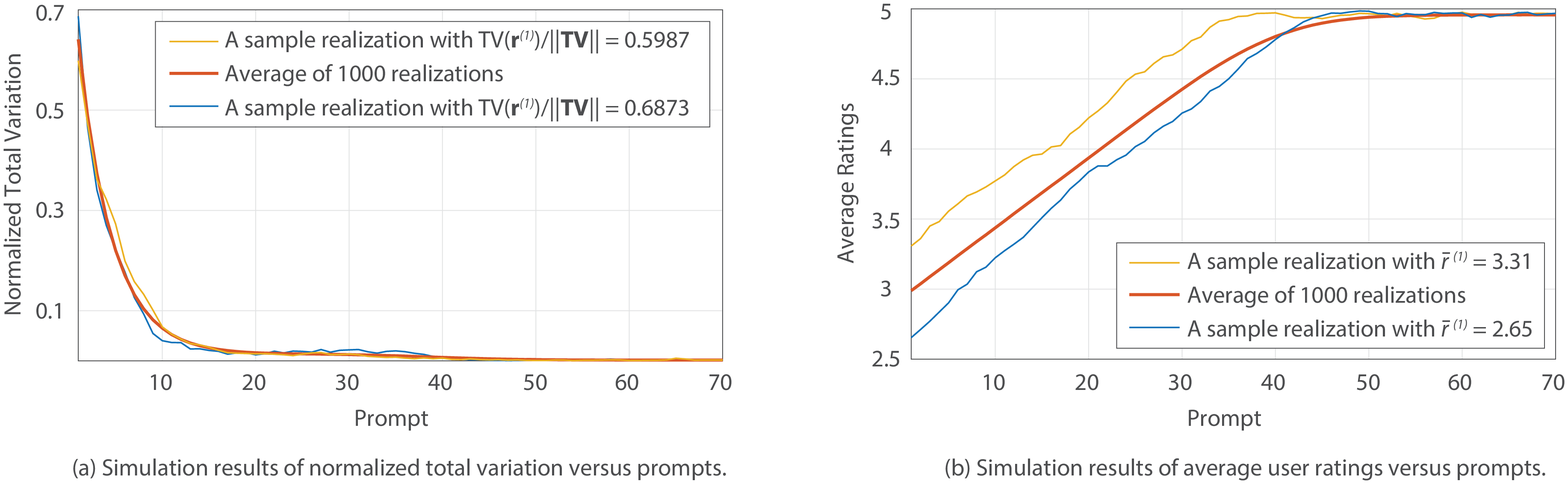}} 
                              
        \caption{Simulation results of normalized total variations and average ratings as functions of prompts. The normalized total variation across the network decreases in (a), while the average ratings saturate to the ceiling in (b).}
        \label{fig:sim}
    \end{figure}

To simulate the proposed Laplacian dynamics-based model described in \eqref{e:model_online}, we run iterations with $\mathbf{r}^{(1)}$ as initialization, where $r_{i}^{(1)}$'s are drawn uniformly at random from the interval $[1,5]$. Figure~\ref{fig:sim}(a) shows two realizations of the total variation \eqref{total_variation} as it evolves over prompts, superimposed to the mean evolution obtained after averaging 1000 independent such realizations. The specific realizations show fluctuations similar to those observed in Figure \ref{fig:tot_var_over_rat}(a), but the trend is diminishing towards zero meaning that ratings converge with temporal evolution. Likewise, Figure \ref{fig:sim}(b) shows two realizations and the mean evolution of the network-wide average ratings $\bar{r}^{(p)}$ in Eq.~\eqref{average} as a function of prompts.

\begin{figure*}[h]
        \centering    
        \includegraphics[width=0.7\linewidth]{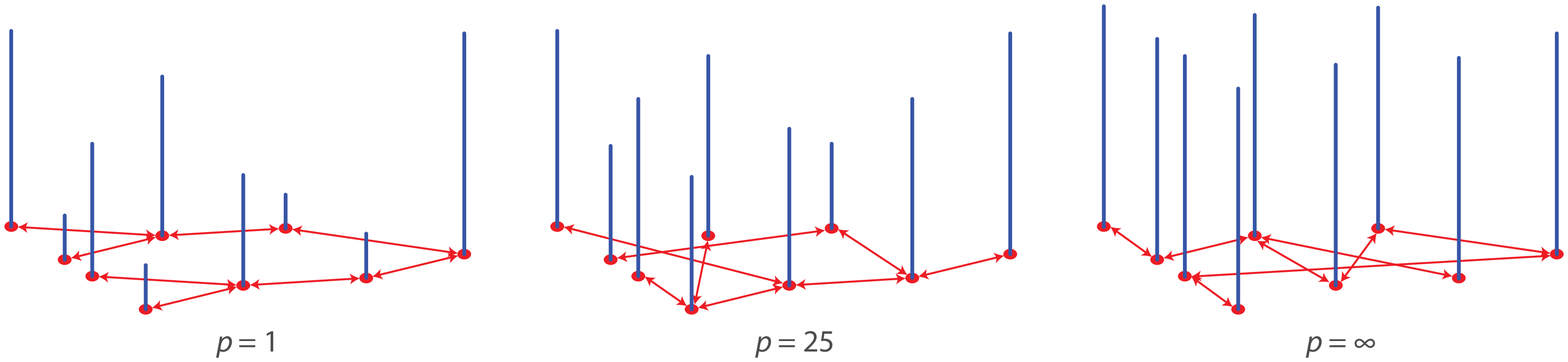}
                              
       \caption{A sample visualization of how the participants' ratings $\mathbf{r}^{(p)}$ (blue bars) change as total variation diminishes and average rating saturates at limiting time (i.e., prompt $p \rightarrow \infty$). The network topology changes in every prompt as the participants randomly interact with each other. A decreasing $\text{TV} (\mathbf{r})$ measure is a direct manifestation of the signal $\mathbf{r}^{(p)}$ being smoother as $p$ increases.}
        \label{fig:converge}
    \end{figure*}

The figures indicate that the simulation results accurately capture the positively skewed effects of interactions in the gradual buildup of speaking skills. While the network model is admittedly simple, the synthesized sample paths qualitatively match the trends in the data. This is further illustrated in Figure~\ref{fig:converge}, where participant-ratings are shown as blue bars demonstrating a decrease in the variation of ratings as the simulation evolves in time, with the average ratings approaching the limiting value $r_{\max}$.

\section{Discussion}

In this paper, we explored how participants' performance ratings progressed as they interacted in online learning communities. In particular, we asked whether the learners' ratings were affected by their interaction patterns and their peers' ratings, and also whether the communities gradually approached homogeneity in speaking performance ratings. We found positive evidence for both of the questions.

Developing competence in speaking skills has its importance across professional and personal settings. There have been some explorations in directional peer tutoring when it comes to developing oral expression in learning a second language --- namely between two students in school or between a family member and a student at home~\citep{duran2016reading,topping2017effective}. However, in the setting of small learning groups, the network-effects in developing one's speaking skills received negligible attention in literature --- a context we presented our insights in. Our novel dataset of six online communities thus allowed us to study the peer-effects in an objective manner. 

We studied performance ratings as graph signals on top of comment-based interaction networks. The emerging field of GSP has received attention from a variety of fields such as Neuroscience and medical imaging, but GSP was not applied to a learning setting previously. Taking a GSP-based prediction approach helped us observe that the learners' ratings had a connection with their interaction patterns \textit{as well as} the rating signals of the peers they interacted with. Both parts of this observation agree with previous findings in other scopes of learning. For instance,~\cite{fidalgo2015using} had shown that there is a direct relation between quantities of interaction and individual performance outcomes. Using Texas Schools Microdata, it was shown that an exogenous change of 1 point in peers’ reading scores raises a student’s own score between 0.15 and 0.4 points \citep{hoxby2000peer}. Our prediction formulation outperformed the baseline relatively by 23\%, thus corroborating the same insights in the context of speaking skill development. 

Moreover, evidence for a gradual buildup of skills towards community-wide homogeneity was not reported in earlier learning community literature, and that is a gap we addressed. Theoretically, this effect has long been anticipated. For instance,~\cite{cowan2004network} simulated diffusion of knowledge and suggested its gradual buildup in networks. Following the collaborative model of learning, this buildup is also reasonable, as repetitive interactions can be anticipated to help participants develop their tacit understandings of speaking skills gradually~\citep{davidson2014boundary}. In our dataset, total variation of the rating signals going down (slope = $-0.04$, $p<0.01$) and the average ratings going up (slope = $0.07$, $p<0.05$) with time provide evidence of skill-ratings building up among interacting participants. To the best of our knowledge, this is the first data-driven evidence for this idea. We further proposed a network diffusion model and used it to simulate the rating buildup process. The model is simple and derived from previous work~\citep{olfati2004consensus}, but we modified the formulation therein to capture the peer-learning effects as observed in our analysis of the data.

It is important to stress that while our dataset is concerned with speaking skill development and therefore our insights are presented in that context, the analysis methods are not limited to speaking-skills only. Any dataset in a learning setting that comes with quantifiable learning outcome measures (e.g., grades, scores, ratings) as well as interaction measures (e.g., comments or other synchronous/asynchronous textual discussion, audio, video, or any suitable Internet-supported media based interaction) can be modeled and analyzed using our formulations.

In this paper, we did not take into account the contents of the comments, which is left as a future work. We treated each comment as an interaction unit. This still turned out to hold information regarding skill-development in the communities, as shown through our mining of the data. Similarly, the exact specifics of the speaking skills were not considered in this paper, such as the use of gestures, vocal variety, friendliness, clarity, speaking speed etc. Rather, we used an average `overall' rating that we assumed to be an objective abstraction of speaking performance or skills. Digging deeper into the comments and connecting them with specific attributes of speaking-skills (e.g., a comment on vocal variation leading to better vocal performance) in the future might help us uncover more insights on the peer-influence characteristics.

In his study on behavior propagation, \cite{centola2010spread} showed that health behavior spreads farther and faster across clustered-lattice networks than across corresponding random networks, since the former structure facilitates social reinforcement through redundant network ties. \cite{valente2012network} reviewed various social intervention approaches for accelerating behavior change. Exploring the effects of various network structures and intervention strategies to maximize skill development are some of our future courses of work. We thus anticipate that the current study will open up many avenues for broader exploration of peer-influenced skill development in the future.

\section*{Data Availability}
\begin{flushleft}
The dataset  and the analysis codes can be found in:\\ 
(Github link, omitted for anonymity)
\end{flushleft}

\bibliographystyle{dcu}

\section*{Author contributions statement}

All the figures are prepared by the authors. Details of contributions have been omitted for anonymity.

\section*{Additional information}
The authors declare no competing financial interest.

\end{document}